# Sr- and Ni-doping in ZnO nanorods synthesized by simple wet chemical method as excellent materials for CO and $CO_2$ gas sensing

Parasharam M. Shirage[1,†], Amit Kumar Rana[1], Yogendra Kumar[1], Somaditya Sen[1], S. G. Leonardi[2] and G. Neri[2,*]

[1]Department of Physics and Centre of Materials Science and Engineering, Indian Institute of Technology Indore, Simrol Campus, Khandwa Road, Indore 453552, India

[2]Department of Engineering, University of Messina, Messina 98166, Italy.

[†]Author for correspondence (E-mail- pmshirage@iiti.ac.in, paras.shirage@gmail.com)

* Email: gneri@unime.it

**Abstract**: In this study, the effect of Sr- and Ni-doping on microstructural, morphological and sensing properties of ZnO nanorods has been investigated. Nanorods with different Sr and Ni loadings were prepared using a simple wet chemical method and characterized by means of scanning electron microscopy (SEM), X-ray diffraction (XRD) and photoluminescence (PL) analysis. XRD data confirmed that Sr- and Ni-doped samples maintainsthe wurtzite hexagonal structure of pure ZnO. However, unlikes Sr, Ni doping modifies the nanorod morphology, increases the surface area (SA) and decreases the ratio of $I_{UV}/I_{green}$ photoluminescence peak to a greater extent. Sensing tests were performed on thick films resistive planar devices for monitoring CO and $CO_2$, as indicators of indoor air quality.The effect of the operating temperature, nature and loading of dopant on the sensibility and selectivity of the fabricated sensors towards these two harmful gases were investigated. The gas sensing characteristics of Ni- and Sr-doped ZnO based sensors showed a remarkable enhancement (i. e. the response increased and shifted towards lower temperature for both gases) compared to ZnO-based one, demonstrating that these ZnO nanostructures are promising to fabricate sensor devices for monitoring indoor air quality.

# 1. Introduction

Metal oxide semiconductor (MOS) gas sensors have been intensively studied for applications in the detection of hazardous pollutant gases[1, 2]. They have many advantageous features respect to other gas sensor devices, such as high sensitivity, low power consumption, low price, quick response and simple structure with high microelectronic processing compatibility. Actually, the research for improving the performance of MOS sensors is addressed towards novel syntheses and/or modification of nanostructured metal oxides with improved sensing characteristics[2].

Among these, ZnO is a well-known material for semiconductor device applications[3]. It has a direct and wide band gap in the near-UV spectral region[4] and has noticeable gas sensing properties[5]. At present, numerous efforts have been made to improve the sensing properties of MOS gas sensors based on ZnO. Low dimensional nanostructures such as nanorods, nanotube, nanoflower, *etc*., showed peculiar characteristics[6-8]. ZnO nanostructures demonstrated novel applications in optoelectronics, sensors, transducers and biomedical sciences[9, 10]. With reduction in size, novel electrical, mechanical, chemical and optical properties are introduced, as a result of surface and quantum confinement effects and are of benefit for developing nano-devices of new generation with high performance.

Doping of base metal oxide with various metallic elements, for example, noble metals, transition metal and metal oxides, had been proven also effective for this scope[11, 12]. Indeed, doping enhances the performance gas sensor via controlling donor density, changing the acid-base properties and varying electronic properties so to change the interaction between gases surrounding the sensor and sensing layer.

In this paper, we investigated ZnO nanorods doped with Sr and Ni. Very few papers regarding Ni-doped ZnO sensors are reported in the literature[13-17], whereas, at the best of our knowledge, only one paper deals instead with sensing properties of Sr-ZnO sensors[18]. Therefore, we decided to evaluate the sensing capability of ZnO nanorods doped with different loading of Sr and Ni synthesized by a simple wet chemical method. The morphology and microstructure of the synthesized materials were first investigated. Then, they were used to fabricate resistive gas sensors for monitoring CO and $CO_2$,. Sensors that measure accurately CO and $CO_2$ concentration in air are highly required for indoor air quality control. CO is a concern in indoor environments where poorly ventilated appliances are present or where outdoor air intakes are located in areas subject to high concentrations of vehicle exhaust[19]. OSHA organization has established a permissible exposure limit for CO of 50 ppm as an 8-hour time-weighted average. $CO_2$ levels can be used as a rough indicator of the effectiveness of ventilation, and excessive population density in a structure. The eight-hour permissible limit for $CO_2$ is 5000 ppm[20]. Then, sensors with promising sensing characteristics to these two target gases are highly desired to be applied in electronic devices for monitoring indoor air quality and used to trigger an alarm, turn on a ventilation fan, or control a heating, ventilating and air conditioning (HVAC) system.

## 2. Experimental details

Pure, Ni- and Sr- doped ZnO nanorods were prepared by simple wet chemical method. High purity chemicals zinc acetate, nickel nitrate, and strontium nitrate (from Alfa Aesar) were used as the raw materials. First, the stoichiometric amount of zinc nitrate, and nickel nitrate or strontium nitrate were dissolved in 100 ml double distilled water and stirred for 30 minutes. Then, under constant stirring, aqueous ammonia solution was added continuously to maintain a pH ~ 11. Finally, a transparent solution is obtained in the case of pure ZnO and Sr-doped, whereas a blue solution is obtained in case of Ni-doped ZnO. In the next step, the solutions were kept at 100°C for 2 hours and then the precipitates were separated and annealed at 250°C (for 2 hrs.). The pure ZnO, Ni- (5% and 10%), and Sr- doped (4% and 8%) samples are named ZO, Ni5ZO, Ni10ZOand Sr4ZO, Sr8ZO respectively.

The morphology of the prepared samples was investigated by field emission scanning electron microscopy (FESEM) using Supra 55 Zeiss microscope. The crystalline structure was verified by Bruker D8 Advance X-ray diffractometer with Cu-K$\alpha$ radiation ($\lambda$=1.54 Å). The average grain size (D) was calculated from the XRD result by using the Scherer formula: D= K $\lambda$/$\beta$ cos$\theta$, where K is the shape factor (~0.9), $\lambda$ is the X-rays wavelength, $\theta$ is the diffraction angle and $\beta$ is the full width at half maximum of the XRD peak. The nitrogen adsorption and desorption isotherms were obtained by Quantumchrome $Q_2$, and the specific surface area and the pore size distribution were calculated by the Brunauer-Emmett-Teller (BET) and the Barrett-Joyner-Halenda (BJH) methods, respectively. The room temperature photoluminescence (PL) measurements were carried out on Dongwoo Optron spectrophotometer. The excitation wavelength was fixed at 325 nm and emission spectra were scanned from 300 to 850 nm.

Sensing tests were carried on resistive sensors having a planar configuration and based on alumina substrates (6 mm × 3 mm) with Pt interdigitated electrodes and a Pt heater located on the back. The devices for electrical and sensing tests were prepared by printing films (~10 μm thick) of the nanorods dispersed in water. A picture of the final device with the Ni10ZO printed film after annealing at 350°C is shown in Fig. 5.

The sensors were then introduced in a Teflon test chamber for the sensing tests. The electrical measurements were carried out over the temperature range from RT to 350°C under a synthetic dry air (20% $O_2$- 80% $N_2$) stream of 100 sccm by collecting the electrical resistance of the sensitive films. An Agilent 34970A multimeter data acquisition unit was used for this purpose, while an Agilent E3632A dual-channel power supplier instrument was employed to bias the built-in heater of the device. Sensing tests were performed by injecting pulses of the analyte from certified bottles. The concentration of the target gas was varied by using mass flow controllers. The gas response is defined as the ratio $R_{air}/R_{gas}$, where $R_{air}$ is the electrical resistance of the sensor in dry air and $R_{gas}$ the resistance at different target gas concentrations.

## 3. Results and discussion

*Characterization*

X-ray diffraction (XRD) analysis was performed to examine the effect of Sr and Ni doping on the crystal structure of the ZnO nanorods. Fig. 1 shows the XRD diffraction pattern of the pure, Sr- and Ni-doped ZnO samples. The diffraction peaks in the pattern of pure ZnO can be indexed to hexagonal wurtzite structured ZnO (space group: P63mc (186); *a* = 0.3253 nm, c = 0.5211 nm) in agreement with JCPDS card for ZnO (JCPDS 036-1451). The high intensity of the peaks demonstrates the good crystallinity of the samples.

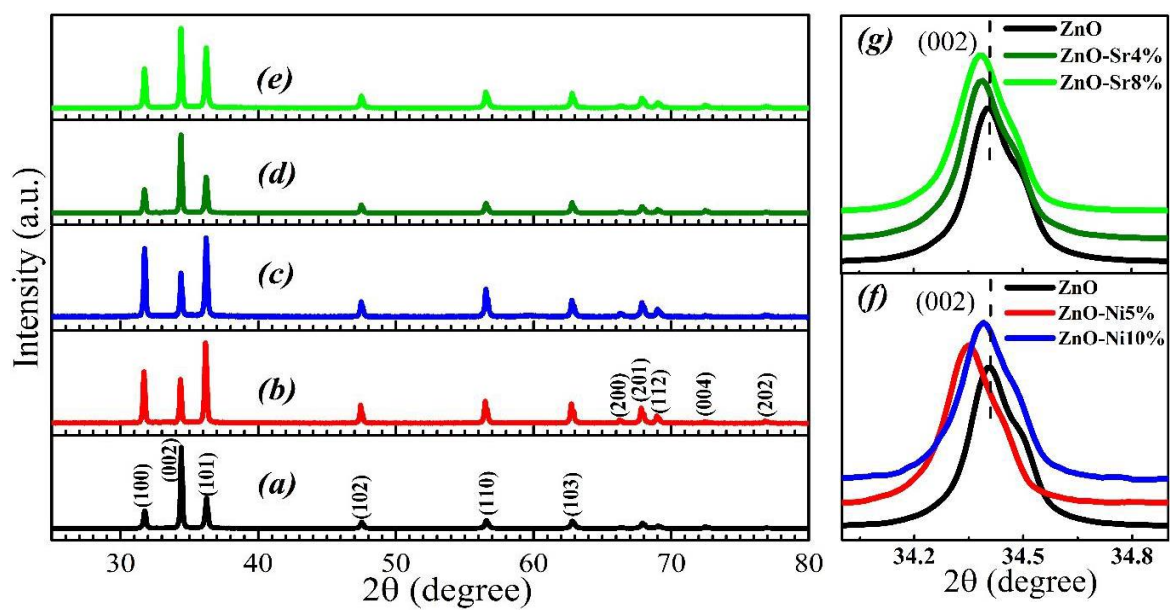

*Figure 1. XRD patterns of the powder sample (a) ZO (b) Ni5ZO (c) Ni10ZO (d) Sr4ZO (e) Sr8ZO. Relative shift in (002) peak of ZnO with respect to (f) Sr-doping and (g) Ni-doping.*

The nanocrystallites of the pure ZnO are preferentially oriented along the *c*-axis, [002] direction. The doping with Sr and Ni change the relative intensity of the diffraction peaks, indicating a modification of the axis orientation, in special way for Ni-doped samples. No characteristic peak of any other possible phases, such as NiO, has been observed, indicating the easy substitution of Ni in ZnO, likely due to similar ionic radius of $Ni^{2+}$ ion (0.68 Å) with $Zn^{2+}$ (0.74 Å)[21]. The (002) peak was used for the calculation of average grain sizes and results for all samples are shown in Table 1. No remarkable difference in the grain size among the different samples was evinced. Instead, a shift of XRD peaks of doped ZnO to lower 2θ value relative to those of undoped ZnO is observed for the incorporation of Ni ions into ZnO. The (002) reflection peak in Ni5ZO show a slight shift towards a lower value relative to undoped ZnO. Further, (100) and (101) planes diffraction peaks are prominent than (002) plane, which indicates the change in the orientation/morphology with increasing Ni doping[6].

| Sample | $D_{(002)}$ (nm) |
|--------|------------------|
| ZO     | 85.8             |
| Ni5ZO  | 85.1             |
| Ni10ZO | 81.5             |
| Sr4ZO  | 84.4             |
| Sr8ZO  | 86.0             |

*Table 1. Average grain sizes calculated from the (002) reflection.*

This expansion of the lattice constants or decrease in 2θ value for Ni5ZO nanostructure hint the strain relaxation is achieved[22]. The calculated values of lattice parameters in the hexagonal crystal are shown in Fig. 2. For higher Ni loading (Ni10ZO sample), we observe a decrease for both *a* and *c* lattice parameters, *i.e.* peak shift towards the higher theta value, due to the Ni-O bond length smaller than Zn-O[15].

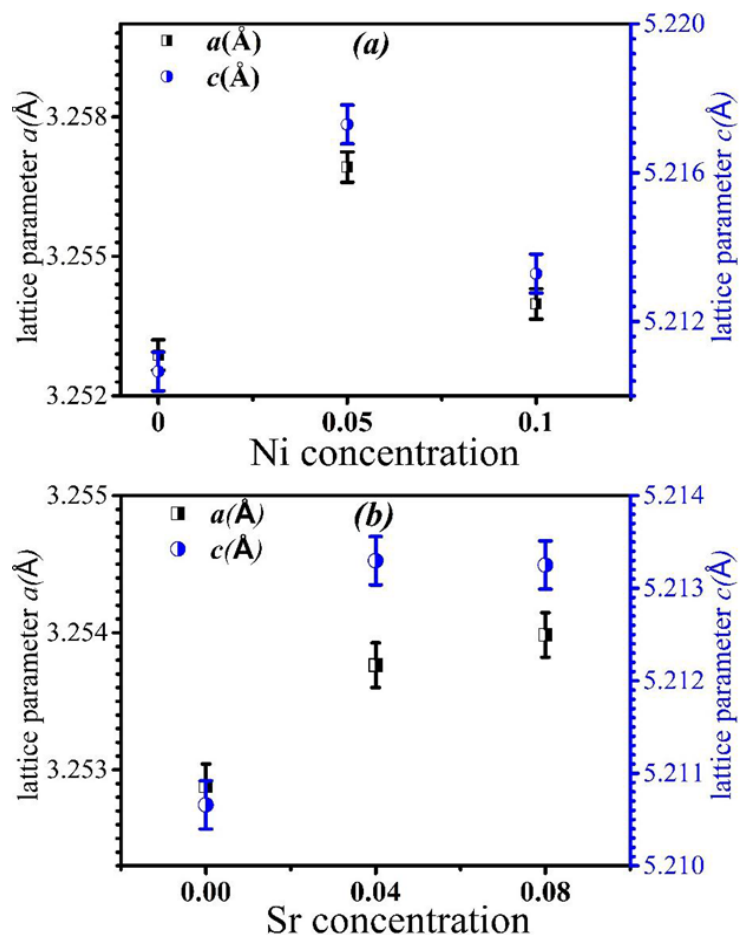

*Figure 2. Variation of lattice parameters a and c with (a) Ni doping and (b) Sr doping in pure ZnO.*

For Sr-doped samples, there is increment in both *a* and *c* parameters because of large ionic radii of $Sr^{2+}$ (1.18 Å) than $Zn^{2+}$. It is also observed that small change in lattice parameter is due the lower solubility of $Sr^{2+}$ in ZnO because of its large ionic radii. A similar result has been also observed by L. Xu *et al*.[23] and Water *et al*.[24] This verifies that Sr atoms are really incorporating into ZnO. In Sr doped samples the lattice orientation is still along *c*-axis as compared to Ni doped samples. Hence, we conclude that Sr doping is maintaining *c*-axis orientation of pure ZnO.

SEM images of the various samples reported in Fig. 3 (*a-e*) reveal that Ni doping causes a change in the nanorods morphology, while Sr doping create pores in nanorods.

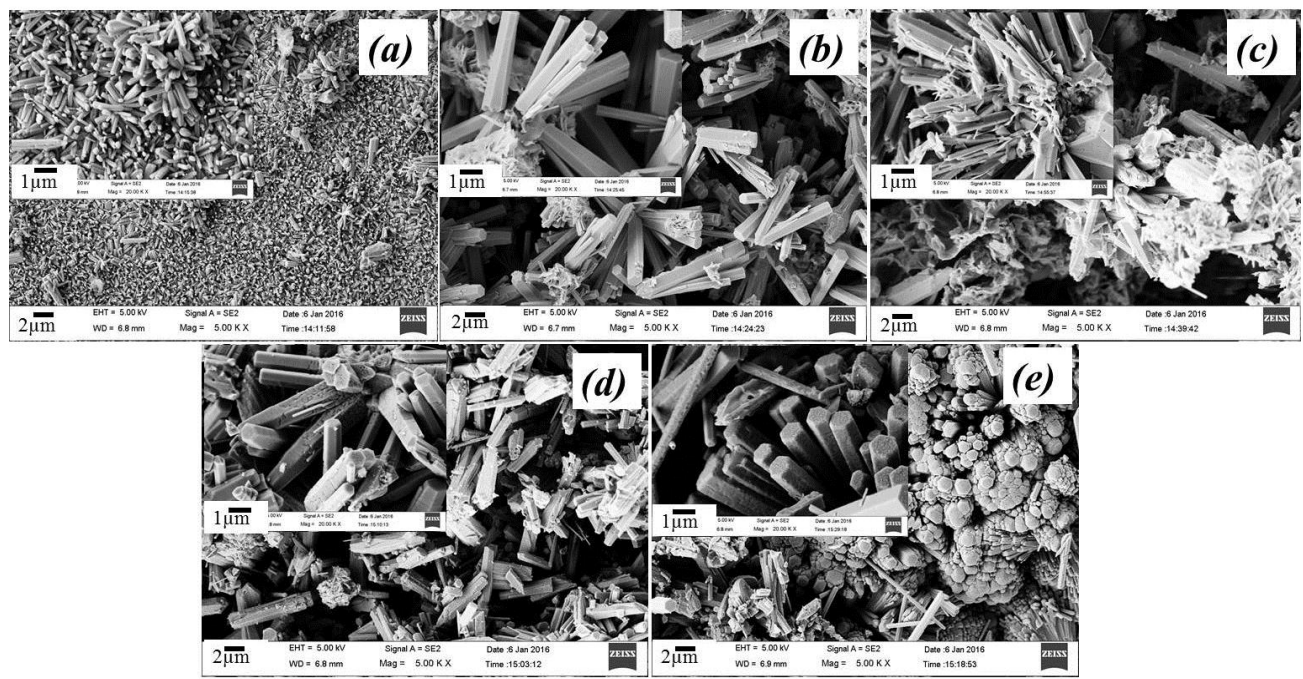

*Figure 3. SEM images of the powder sample (a) ZO (b) Ni5ZO (c) Ni10ZO (d) Sr4ZO (e) Sr8ZO (inset images at 20 KX).*

The growth mechanism of ZnO was already discussed elsewhere[6]. In pure ZnO average length and diameter of nanorods are about 1-2 µm and 200-400 nm respectively. In case of 5% Ni doping the average length and diameter of nanorods increase to 2-4 µm and 600-800 nm respectively. Again with higher Ni concentration there is no major change in length and diameter but with 10% Ni doping there is a change in morphology from rods into flakes, which is due to the incorporation of Ni ions in ZnO structure. In Sr 4% doping the range of length and diameter are 4-6 µm and 600-800 nm respectively while in Sr 8% there is no significant change in dimension and morphology of nanorods as compare to Ni doping ZnO. It has been observed that in higher Sr doping there is formation of bunches of rods as compared to lower doping where we are getting separate nanorods. Another interesting fact observed is the formation of pores in Sr-ZnO nanorods with increasing Sr concentration, due to large difference in ionic radii between the host matrix and the dopant.

In order to evaluate the textural effects induced by the addition of dopants, the nitrogen adsorption-desorption isotherm and the corresponding Barrett–Joyner–Halenda (BJH) pore size distribution plot (see inset) of all samples are shown in Figure S1. Using the BJH method and the desorption branch of the nitrogen isotherm, the BET surface area (SA) have been calculated and are shown in Table 1. Ni doping increase strongly the surface area (more than tenfold), unlikely Sr doping which lead instead to a decrease of SA. Interestingly, the pore size distribution of pure and Ni-doped samples revealed a prominent peak at very small pore radius, whereas the opposite was found for the Sr-doped samples. On the basis of SEM and XRD characterizations, it appears that the SA enhancement of SA is mainly linked to change of nanorod morphology, rather than a change in their size. The lower SA area of Ni-doped samples could be instead due to the increase of pore radius with Sr doping.

Fig. 4 shows the PL spectra of Sr- and Ni-doped ZnO nanorods with different doping concentrations. Two peaks are observed in all samples, an ultraviolet (UV) emission (around 385 nm), attributed to near band-edge emission - namely, the recombination of free excitons through an exciton-exciton collision process, and a broad green emission peak (around 580 nm) arising from the intrinsic defects in the ZnO nanorods such as the Zn interstitials and the oxygen vacancies-related donor defects[17, 25-27].

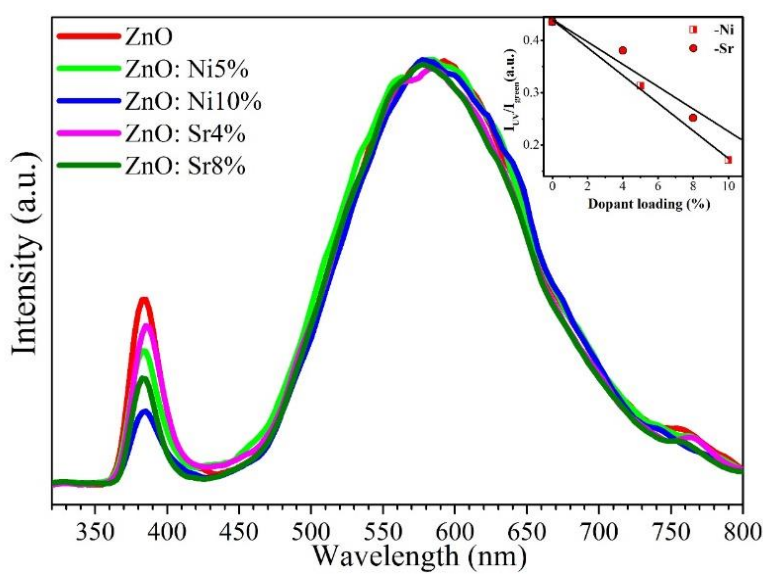

*Figure 4. PL spectra at room temperature with excitation energy 325 nm in the range 330 nm to 800 nm. Inset shows the $I_{UV}/I_{green}$ ratio as a function of dopant loading.*

UV emission strongly relate to crystallite quality of ZnO. As clearly shown in Fig. 4, where the PL curves have been normalized at the maximum of green emission peak at 580 nm, a decrease of UV emission follows the increase in Sr- and Ni-doping. By doping ZnO with high amount of Sr and Ni, these atoms go into lattice sites substituting for Zn atom causing lattice distortion and leading to decrease of UV emission peak. Green emission, centered at around 580 nm, is attributed to ZnO surface defect in the sample. Sr- and Ni-doping can also introduce impurity defects, such as interstitial and substitution defects, which contribute to the increase of visible (green) emission in PL, contributing to decrease the $I_{UV}/I_{green}$ PL peak ratio. Has been reported in literature that decreasing $I_{UV}/I_{green}$ PL peak ratio (i.e. corresponding to an increase in concentration of oxygen vacancies) can improve the sensing properties[28]. In fact with increasing the concentration of oxygen vacancies as electron donors in $ZnO_{1-x}$ lattice, the amount of active absorbed oxygen species increases. From data reported in the inset of Fig. 4 it seems that Ni doping decreases the ratio of $I_{UV}/I_{green}$ photoluminescence peak to a greater extent than Sr.

*CO and $CO_2$ sensing tests*

CO and $CO_2$ are major air pollutants whose concentration has to be controlled as per international indoor air quality standards at industrial and public work places. Then, there is a high interest to fabricate efficient CO and $CO_2$ sensors. Resistive sensors having a planar configuration (Fig. 5) have been fabricated by printing thick films of the synthesized samples on the Pt interdigitated electrodes of an alumina support.

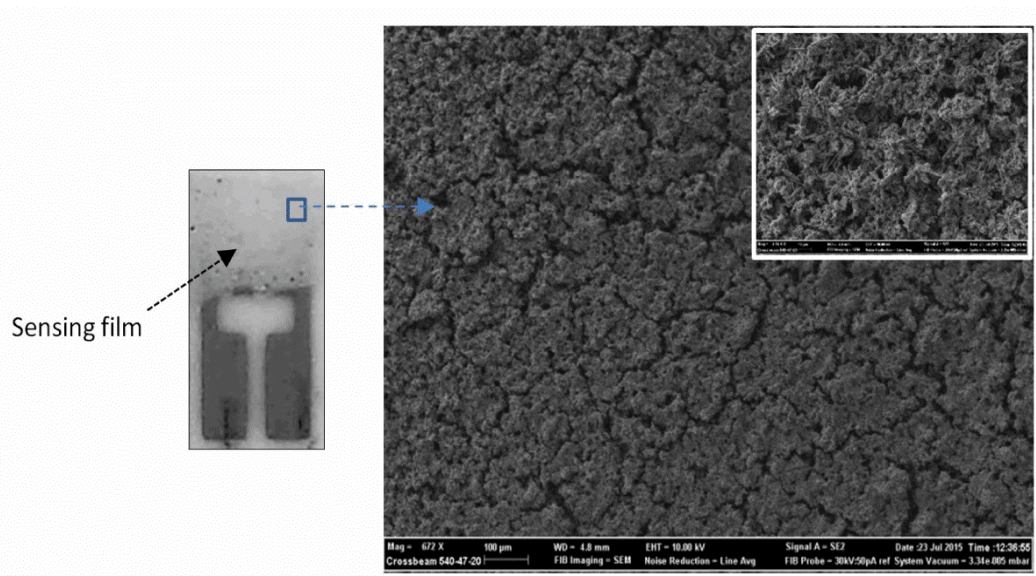

*Figure 5. Photograph of the sensor device and micrograph of the sensing film surface. Inset shows a high magnification of the highly porous sensing film surface.*

The surface of the sensing layer shows a highly porous structure. Before sensing tests all the sensors prepared were heated up to 350°C and left to stabilize for a time of about 2 hours in order to stabilize the printed films. In Figure 6 is shown a panoramic view of the dynamic response of some samples investigated towards CO and $CO_2$ at different temperatures. The response to a decrease of oxygen, from 20% to 10% (v/v), is also shown. Operating temperature was found to plays an important role in gas response of the investigated sensor. In the temperature range between 200°C and 350°C all samples show responses to CO and $CO_2$. In accordance with n-type semiconductor behavior of the samples, after pulses at different concentration of CO and $CO_2$ in synthetic air, a decrease of the electric resistance is observed. Indeed, when a reducing gas such as CO is added, the interaction of this gas with the surface chemisorbed oxygen, $O^{2-}$, can take place. The reducing gas readily releasing electrons back to the conduction band, according eq. (1), and the electrical resistance of the semiconductor decreases.

$$CO + O^{2-}_{(ads)} \rightarrow CO_2 (g) + 2e^- \quad \text{-------------------- (1)}$$

Different sensing mechanisms are responsible of the electrical behavior of ZnO-based materials towards $CO_2$. Among these, the reactions of $CO_2$ with adsorbed $OH^-$ on the surface to form carbonates and hydroxycarbonates are the most important. These intermediate species react with adsorbed oxygen on the ZnO surface, releasing electrons and, consequently decreasing the resistance. Furthermore, at removal of the target gas and subsequent exposure again to synthetic air, all the samples are able to recover the baseline. For example, even at the low operating temperature of 200°C, the dynamic response shown by the Ni-ZnO sensor is rather fast ensuring rapid response times of about 15 seconds and recovery less than 90 seconds.

It is noteworthy that at lower temperature (200°C), the pure ZnO and Sr doped sensors exposed to CO pulses present an inversion of the response (see Fig. S2 in Supporting Information). In the presence of $O_2$, the $O^-_{ads}$ species becomes dominant at an operation temperature of 100°C-200°C. It is generally accepted that CO interacts with $O^-_{ads}$ to form carboxylate and bi-dententate carbonate (*equ.* 2 & 3), which helps in releasing electrons in the conduction band and thus decreasing conductance (i.e. increase in resistance) of Sr-doped ZnO nanorods[29].

$$CO + O^-_{ads} \leftrightarrow CO^-_{2,ads} \leftrightarrow CO_2(g) + e^- \quad\quad\quad (2)$$

$$CO + 2O^-_{ads} \leftrightarrow CO_3^{2-} \leftrightarrow CO_2(g) + O^-_{ads} + e^- \leftrightarrow CO_2(g) + 1/2 O_2 + 2e^- \quad (3)$$

Besides this the unusual sensing behavior can be ascribed to the chemical reaction of CO molecule by the adsorption to the grain boundary of oxygen gas molecules which plays the key role in physical barrier for carrier's movement and increase the resistance[30]. The inversion of response indicate further that, at low temperature, the pure ZnO and Sr doped sensors behave as p-type semiconductors. In these materials, majority carriers are holes, so the electron released in equations (2) and (3) contribute to decrease their concentration and consequently the conductance decreases (resistance increases). Due to this fact, the response of the cited sensors is deteriorated. Instead, Ni-doped samples work well also at lower temperature (up to 150°C for 10% Ni doped), showing no response inversion and are able to recover the baseline although with longer times.

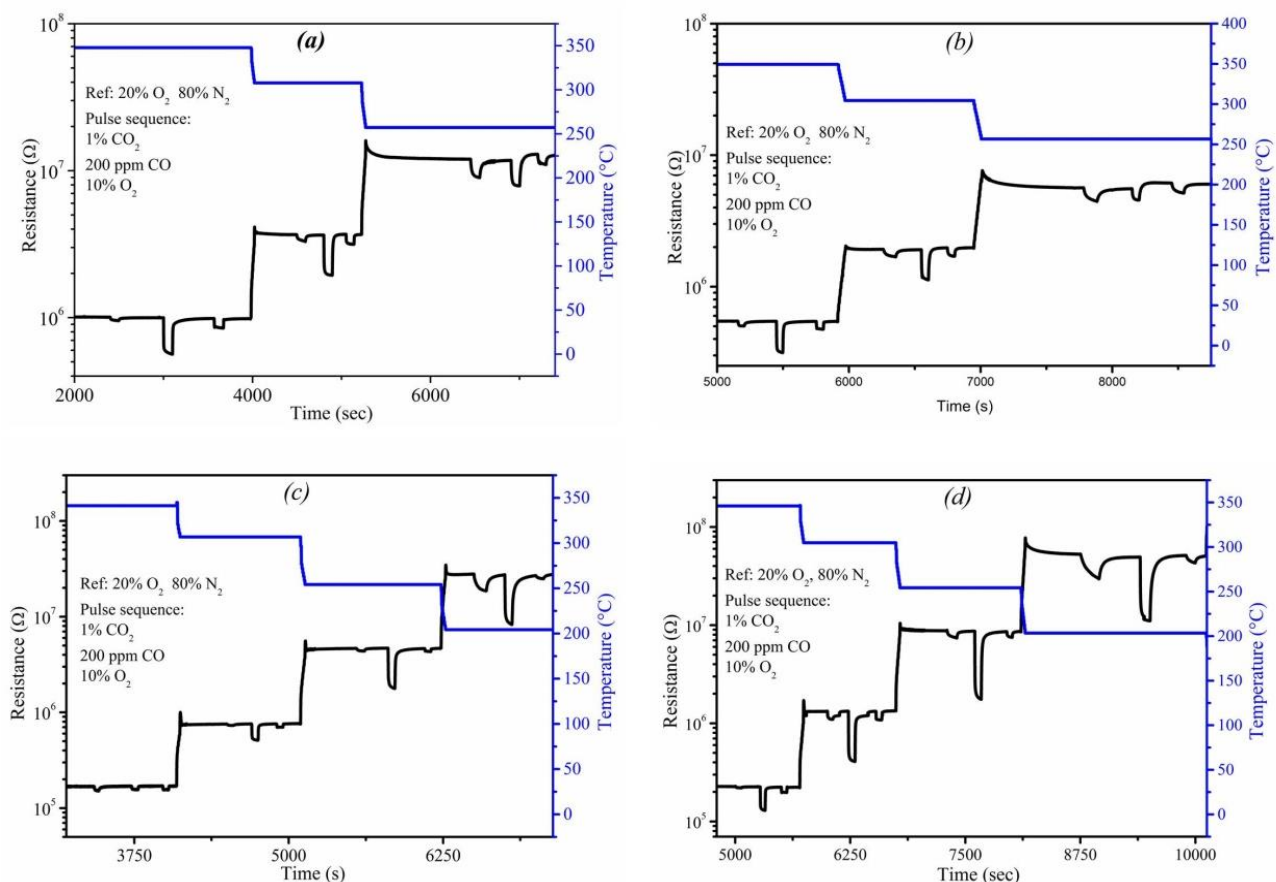

*Figure 6. Dynamic responses of sensors: a) ZO; b) Sr8ZO; c) Ni10ZO; d) Ni5ZO.*

Figure 7 summarizes the responses recorded at different temperatures for pure and doped sensors when exposed to 1% $CO_2$, 200 ppm CO and 10% variation of $O_2$ concentration, respectively. At all temperatures, the sensors are little influenced by oxygen variation. For the two target gases, at temperature higher than 250°C, all sensors exhibit greater response to CO than $CO_2$. Examining the trend of responses to target gases, it is observed that, for Ni-doped sensors, the response to CO is higher with respect the undoped and Sr-doped ZnO sensors and increase with decreasing the temperature. At the temperature of 200-250°C, their response is more than three times than that observed for the ZnO sensor. Hence, doping ZnO with Ni allows a remarkable enhancement of the gas sensing characteristics towards CO.

Vice-versa, Sr-doped sensors display a CO sensing behavior similar to undoped ZnO sensor. In fact, as concerns CO gas, these sensors show a maximum of the response above 350°C, while a drastic loss of response is registered below 250°C. The possible reason for higher CO gas sensing response above 350°C is assumed to higher thermal energy with increase in operating temperature which helps to overcome the activation energy barrier to the reaction and a significant increase in the electron concentration for the sensing reaction to happen[31]. It has been reported that the stable oxygen ions were $O_2^-$ below 100°C, $O^-$ between 100°C and 300°C and $O^{2-}$ above 300°C. The relevant surface reactions can be listed as, [32]

$$2CO + O_2^- \rightarrow 2CO_2 + e^- \text{--------------- (4)}$$

$$CO + O^- \rightarrow CO_2 + e^- \text{--------------- (5)}$$

$$CO + O^{2-} \rightarrow CO_2 + 2e^- \text{--------------- (6)}$$

Combining aforementioned facts and corresponding surface reactions, *equ.* (6) turns out as most desirable for the gas sensing, because the reaction could release more number of electrons to follow a generalized sensing pattern and increase in the gas response. This is only possible when the operating temperature is maintained above 300°C but less than 400°C above which possibly degradation of doped ZnO causes decrease in CO response. In accordance to the equations it can also be explainable that the drastic loss of response for CO gas below 250°C is due to catalytic oxidation of CO by $O_2^-$ (*equ.*(4))[31].

Instead, the response to $CO_2$ is negligible at temperature higher than 250°C, and increases with decreasing the temperature. This opposite trend, indicate that Sr-doped ZnO sensors have the potential as $CO_2$ selective sensors at low operating temperatures.

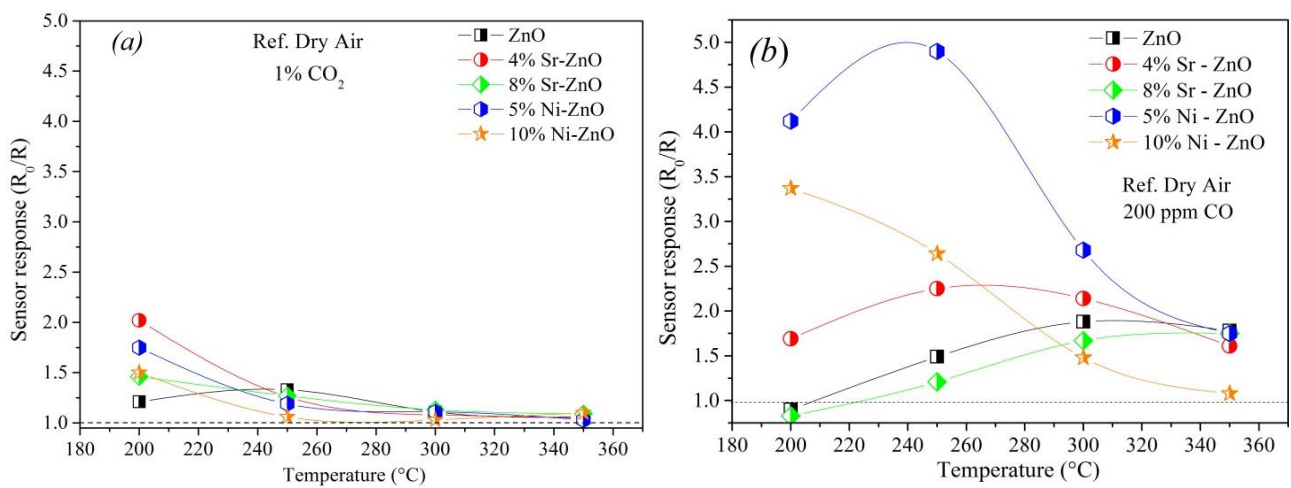

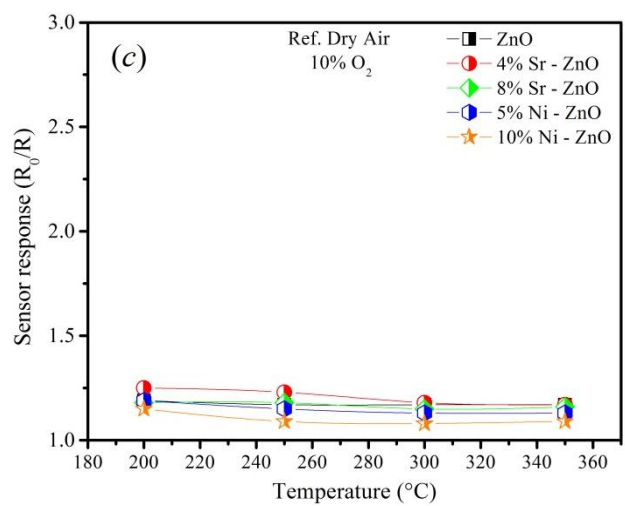

*Figure 7. Sensor responses a) 1% $CO_2$; b) 200 ppm CO; c) 10% $O_2$ at different temperatures.*

The dopant load has a great effect on the response against the tested gases. As shown in Figure 7, the 5% Ni doped sensor shows higher response than the sample with higher Ni loading. This agrees with the reports of other authors. For example, Wang and co-workers[33] reported that the response to $C_2H_2$ has been greatly enhanced by Ni doping ZnO at the optimal doping concentration of 5%. The samples with the excess Ni concentration deteriorate with worse response.

Ni5ZO sensor was also tested at different low CO concentrations (5-80 ppm) at the best operating temperature of 250°C. The dynamic responses and related calibration curve of the sensor are reported in Figure 8. The sensor response shows a linear behavior in the log-log plot with CO concentration. The good response allows to reaching a lower detection limit of less than 2 ppm.

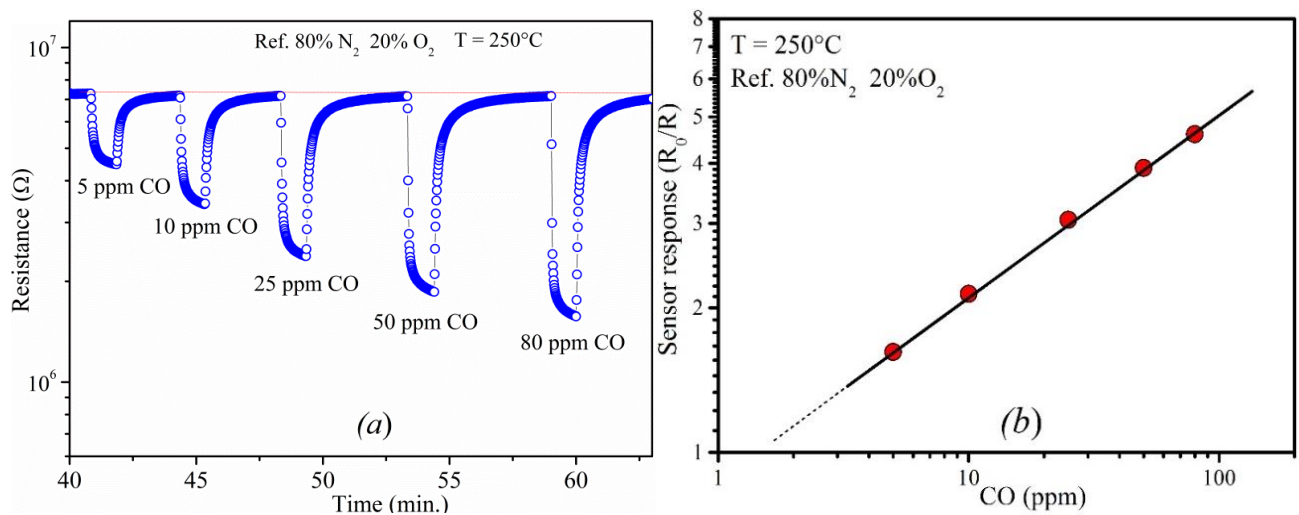

*Figure 8. a) Dynamic response towards different CO concentrations and b) calibration curve at 250°C related to Ni5ZO sensor.*

The same finding has been observed for the $CO_2$, being the Sr4ZO sensor, with an intermediate loading of Sr, the most responsive. In Figure 9 is shown the dynamic response to pulses of different $CO_2$ concentration, in the range of 0.25-2% in volume, at a temperature of 250°C. In these operative conditions the sensor is able to detect $CO_2$ with fast dynamic response. The response of Sr4ZO sensor to both gases at different temperatures is also shown, evidencing as the selectivity towards $CO_2$ for this sensor is maximized at the lower temperature tested (200°C).

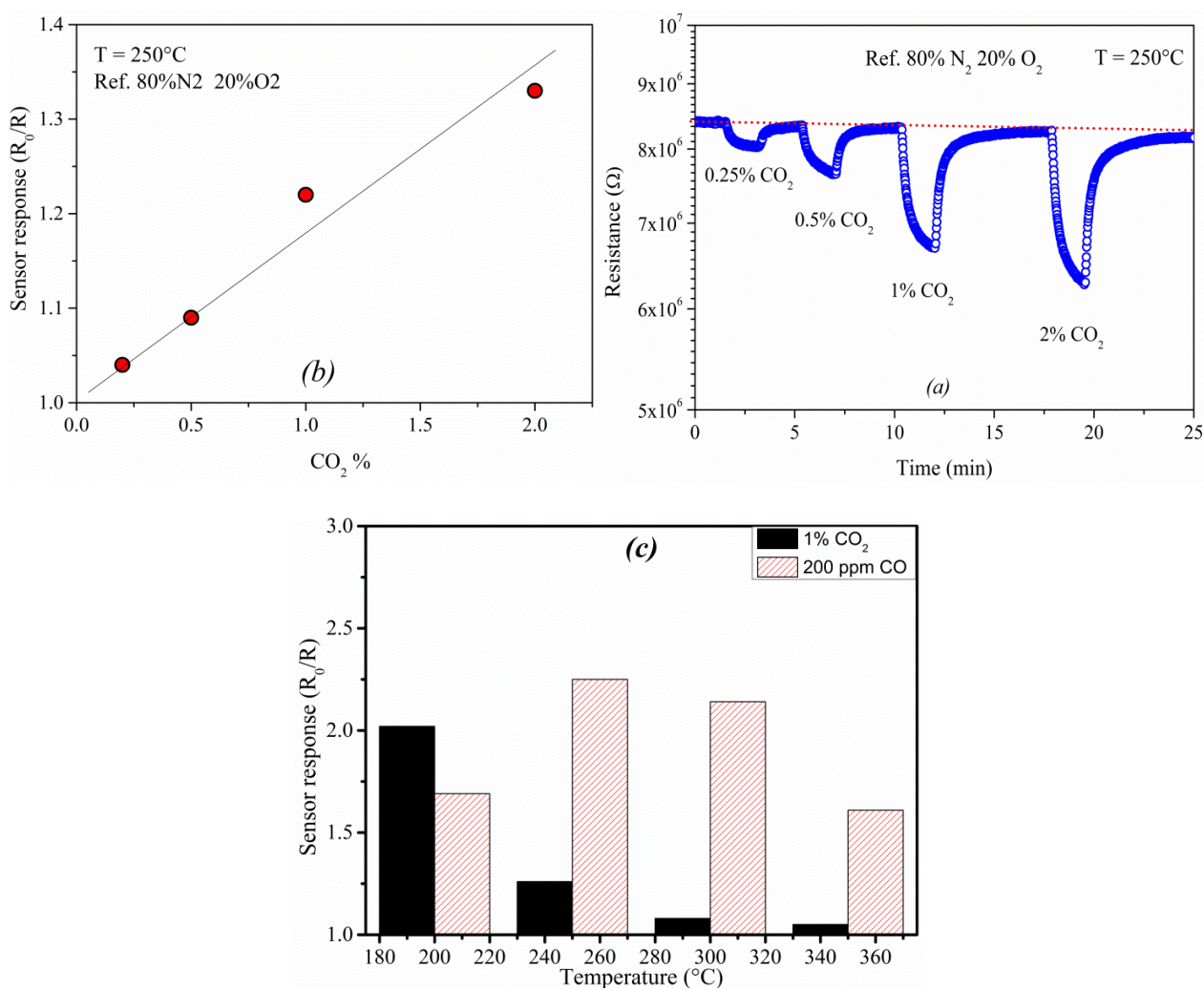

*Figure 9. a) Dynamic response towards different $CO_2$ concentrations and b) calibration curve at 250°C related to Sr4ZO sensor. c) Comparison of the response of Sr4ZO sensor to CO and $CO_2$, at different temperatures.*

Overall results indicated that the Ni5ZO sensor, operating at the temperature of 250°C, is highly sensitive to CO and inherently selective with respect to $CO_2$. As regards the Sr4ZO sensor, although decreasing the temperature the response to CO decrease and that to $CO_2$ increases, we cannot achieve the required selectivity towards this latter specie. In this case, statistical and signal processing techniques are necessary, for example, developing adaptive searching algorithms to retrieve each gas concentrations with improved precision, training data set evaluated by the two-sensor array[34].

A comparison with the sensing performance of some recent CO sensors based on ZnO nanorods reported in the recent literature is shown in Table 2. It can be noted the higher response of our sensor compared to others where ZnO nanorods are doped with different dopants. The search has produced instead no literature data about $CO_2$ sensing using ZnO nanorods-based sensors.

| *Sensing material* | *Temperature (°C)* | *CO (ppm)* | *Response ($R_a/R_g$)* | *Response/ppm × 100* | *Ref.* |
|---|---|---|---|---|---|
| **Ni-ZnO nanorods** | **250** | **200** | **5** | **2.5** | **This work** |
| Ni-ZnO hexagonal plates | 300 | 300 | 2.5 | 0.83 | 35 |
| Co-ZnO electrodeposited nanorods Nanotubes | 350 | 150 | 1.3 | 0.86 | 36 |
| Au/ZnO nanorods | 300 | 200 | 3 | 1.5 | 37 |
| Al-ZnO nanorods | 350 | 100 | 1.7 | 1.7 | 38 |
| Ga-ZnO nanorods | 75 | 200 | 1.1 | 0.55 | 39 |
| CuO-ZnO nanorods | 300 | 200 | 3.5 | 1.75 | 40 |

*Table 2. Comparison of ZnO nanorods-based CO sensors.*

*Sensing response-samples characteristics relationships*

The above reported results demonstrate that the gas sensing characteristics of Ni- and Sr-doped ZnO based sensors towards CO and $CO_2$ are remarkably enhanced compared to ZnO-based one. Indeed, both Sr and Ni increase the response and also shift the maximum response at lower temperatures. An attempt is here made to correlate this behavior with the morphological and microstructural characteristics of the materials used as sensing layers.

Characterization results give indication of the presence of a higher amount of crystal defects in the doped samples. The key role of crystal defects in gas sensing is highly accepted. Hu et al. proposed that the high sensing properties towards of transition-metal doped ZnO nanorods are related to a higher content of donor related defects and a lower content of acceptor related defects, because the donors would provide electrons for the adsorbed oxygen to produce the active ionosorbed oxygen[41]. PL data here reported show a decrease of $I_{UV}/I_{green}$ PL peak ratio with increase of Sr- and Ni-loading, suggesting an increase of the concentration of oxygen vacancies, which can improve the sensing properties[42]. However, in our case, this factor cannot fully explain the decrease of response found on the Sr- and Ni- doped sensors with the higher dopant loading. This means that other factors, such as non-stoichiometry, lattice distortion and smaller grain size, are involved. On the basis of XRD characterization we can exclude the effects due to a decrease of particle size and to the formation of local p–n junction because in our case all $Sr^{2+}$ and $Ni^{2+}$ ions substitute the $Zn^{2+}$ atoms in ZnO lattice and no SrO or NiO phase is found[13]. Also, the sensing characteristics reported for both Ni- and Sr-doped ZnO sensors do not correlate in any way with the surface area of the corresponding samples.

In case of CO response on Ni-ZnO sensors, a good correlation has been found taking into account the lattice distortion (see Fig. 10). Lattice distortion is advantageous for creating defects which are the sites for the interaction between test gas molecule and sensor surface[43].

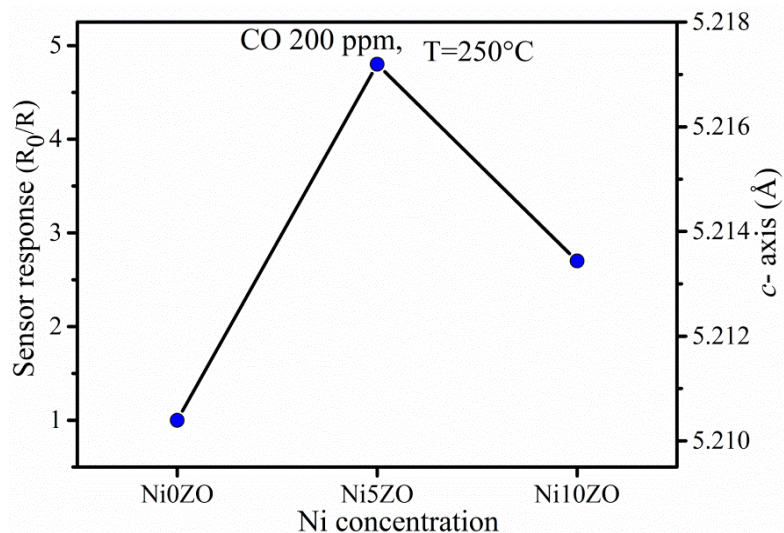

*Figure 10. Sensor response as a function of lattice distortion for Ni-doped ZnO.*

For Sr-doped sensor, the decrease of operative temperature leads instead to a decrease of CO response and correspondingly to an enhancement of the response to $CO_2$. As suggested by us for Ca-doped ZnO sensors, $CO_2$ sensing follows a mechanism relying on the adsorption of $CO_2$ on the sensing layer[44, 45]. On this basis, the $CO_2$ response enhancement for Sr-doped ZnO could be related to the basicity of the dopant. In fact, as an acid molecule, the interaction of $CO_2$ with a solid material to form surface carbonates and hydroxycarbonates is favored if the surface basicity is increased. These intermediate species react subsequently with adsorbed oxygen species releasing electrons and consequently changing the resistance. Then, it seems that the improvement of the sensor response may be attributed to the higher adsorption of $CO_2$ provided by Sr-doped surface which is able to adsorb more strongly the acid $CO_2$ gas molecules compared to a undoped and Ni-doped ZnO surface.

## 4. Conclusion

Sr- and Ni-doped ZnO nanorods were synthesized by simple wet chemical method. Characterization measurements revealed that they maintain the wurtzite hexagonal structure of pure ZnO. However, the decrease of $I_{UV}/I_{green}$ photoluminescence peak ratio and XRD data indicate the presence of lattice distortion, due to the successfully incorporation of dopants within the crystalline structure of ZnO.

The gas sensing characteristics of the chemo-resistive sensors developed with Sr- and Ni-doped ZnO nanorods as sensing layer towards CO and $CO_2$ as target gases were found enhanced with respect to undoped ZnO-based sensor. In the specific, Ni-doped sensors exhibited high response to low concentration of CO in air. The enhanced CO response was attributed to the increased lattice distortion introduced by Ni which favors the interaction between CO gas molecules and the sensor surface. The $CO_2$ response enhancement for Sr-doped ZnO could be instead related to the basicity of the dopant.

On the basis of this report, these simple doped ZnO nanostructures are promising to be applied in electronic devices for monitoring indoor air quality.

## Acknowledgments

This work was supported by the Department of Science and Technology (SERB-DST), India by awarding a prestigious 'Ramanujan Fellowship' (SR/S2/RJN-121/2012) to the PMS. PMS is thankful to Prof. Pradeep Mathur, Director, IIT Indore, for encouraging the research and providing the necessary facilities. We are also thankful SIC IIT Indore for providing the characterization facilities and Dr. Vipul Singh, IIT Indore for his help to do PL measurements.